\documentclass[aps,prl,twocolumn,showpacs,superscriptaddress]{revtex4}
\usepackage{graphicx}
\usepackage{amssymb}
\usepackage{amsmath}
\usepackage{bm}

\usepackage{color}

\newcommand{\ket}[1]{|{#1}\rangle}
\newcommand{\bra}[1]{\langle{#1}|}

\newcommand{\Heff}{H_\textrm{eff}} 
\newcommand{\HeffT}{H_{\textrm{eff},T}} 

\begin{document}

\title{Bulk--Boundary Correspondence for Chiral Symmetric Quantum Walks}
\author{J\'anos K. Asb\'oth}
\affiliation{
Institute for Solid State Physics and Optics, 
Wigner Research Centre, Hungarian Academy of Sciences, 
H-1525 Budapest P.O. Box 49, Hungary}
\author{Hideaki Obuse}
\affiliation{
Department of Applied Physics, Hokkaido University, Sapporo 060-8628, 
Japan}
\date{4th September, 2013}
\begin{abstract}
Discrete-time quantum walks (DTQW) have topological phases that are
richer than those of time-independent lattice Hamiltonians. Even the
basic symmetries, on which the standard classification of topological
insulators hinges, have not yet been properly defined for quantum
walks. We introduce the key tool of timeframes, i.e., we describe a
DTQW by the ensemble of time-shifted unitary timestep operators
belonging to the walk. This gives us a way to consistently define
chiral symmetry (CS) for DTQW's.  We show that CS can be ensured by
using an ``inversion symmetric'' pulse sequence. For one-dimensional
DTQW's with CS, we identify the bulk $\mathbb{Z} \times \mathbb{Z}$
topological invariant that controls the number of topologically
protected 0 and $\pi$ energy edge states at the interfaces between
different domains, and give simple formulas for these invariants. We
illustrate this bulk--boundary correspondence for DTQW's on the
example of the ``4-step quantum walk'', where tuning CS and
particle-hole symmetry realizes edge states in various symmetry
classes.
\end{abstract}

\maketitle

The realization that band insulators can have nontrivial topological
properties which determine the low-energy physics at their boundary
has been a rich source of new physics in the last decade.  The general
theory of topological insulators and superconductors \cite{rmp_kane,
  rmp_zhang} classifies gapped Hamiltonians according to their
dimension and their symmetries \cite{schnyder_tenfold}.  As very few
real-life materials are topological insulators, there is a strong push
to develop model systems, ``artificial materials'', that simulate
topological phases \cite{sengstock_shaken}.  One of the promising
approaches is to use Discrete-time quantum walks
(DTQW)\cite{aharonov93, kempe_2003,ambainis_2003,gabris_prl}, which
can simulate topological insulators from all symmetry classes in 1D
and 2D \cite{kitagawa_exploring, kitagawa_introduction,
  obuse_delocalization}.

DTQW's with particle--hole symmetry (PHS) go beyond simulating
topological insulating Hamiltonians: they have topological phases with
no counterpart in standard solid-state setups. In 1D DTQW's with PHS,
edge states, ``Majorana modes'' can have two protected quasienergies:
$\varepsilon=0$ or $\pi$ (time is measured in units of the timestep
and $\hbar=1$). Building on the results for periodically driven
systems \cite{akhmerov_majorana}, one of us has defined the
corresponding $\mathbb{Z}_2 \times \mathbb{Z}_2$ topological
invariant\cite{asboth_prb}. Both $0$ and $\pi$ energy Majorana edge
states have been experimentally observed in a quantum walk
\cite{kitagawa_observation}.

The situation of chiral symmetry (CS) of DTQW's is much less
clear. Even for the simplest 1-dimensional DTQW, it is disputed
whether it even has CS \cite{kitagawa_exploring} or not
\cite{asboth_prb}. Although it is expected that CS should imply a
$\mathbb{Z}\times\mathbb{Z}$ bulk topological invariant, this has not
yet been found for DTQW's.  As opposed to the case of PHS, there is
also not much to draw on from periodically driven systems.  What
DTQW's have CS? How can the bulk ``winding number'' be expressed for
DTQW's with CS? These are the problems we tackle in this paper.

A DTQW concerns the dynamics of a particle, ``walker'', whose
wavefunction is given by a vector, $\ket{\Psi} = \sum_{x=1}^{N}
\sum_{s=-1,1} \Psi(x,s) \ket{x,s}$.  Here, $x=1,\ldots,N$ is the
discrete position, and $s=\pm1$ indexes the two orthogonal internal
states of the walker, the ``coin eigenstates'', which we also refer to
as ``spin''.  The dynamics, instead of given by a time-independent
Hamiltonian, is realized using a periodic sequence of alternating
``step'' and ``coin rotation'' operations. 

The step operations are translations of the particle by one lattice
site depending on the value of the ``coin'', the $z$-component of its
spin. These are described by unitary operators $S_s$, where $s$ is
either $+$ or $-$, and
\begin{align} 
\label{eq:s_pm_def}
S_\pm \!&=\! \sum_{x=1}^{N} \!\big( \ket{x\!\pm\! 1}\bra{x} \!\otimes\! \ket{\pm 1}\bra{\pm 1} + 
\ket{x}\bra{x} \!\otimes\! \ket{\mp 1}\bra{\mp 1} \big).
\end{align} 
For simplicity, we take periodic boundary conditions.  

Between each two steps, a site-dependent local ``coin rotation'' $R_j$ on
the walker's internal state is performed. We consider
\begin{align}
\label{eq:r_chi_theta_def}
R_j &= \sum_{x=1}^N \ket{x}\bra{x} \otimes R(\chi_j(x),\theta_j(x));\\
R(\chi,\theta) &= 
\begin{pmatrix}
\cos \theta - i \sin \theta \sin \chi & -\sin \theta \cos\chi\\
\sin \theta \cos \chi & \cos \theta + i \sin \theta \sin \chi 
\end{pmatrix}\\
 &= \exp[-i \theta (\cos(\chi) \sigma_y+\sin(\chi) \sigma_z)]
\end{align}
This allows breaking PHS via the angle $\chi$ 
\cite{kitagawa_exploring}. 
Details of how the local operations $R_j$ are performed do not
influence the DTQW, all the information about them is
summarized in the corresponding unitaries $R_j$.

One period of the DTQW is defined by $\ket{\Psi(t+1)} = U_0
\ket{\Psi(t)}$, for $t\in \mathbb{Z}$. Here the unitary timestep
(Floquet) operator is composed of $2M$ successive pulses,
\begin{align} 
U_0 &= S_{M} R_{M} S_{M-1} R_{M-1}\ldots S_1 R_1.
\label{eq:U_def}
\end{align}
A period has to include an equal number of $S_+$ and $S_-$ pulses,
otherwise timestep operator has quasienergy winding
\cite{kitagawa_periodic}, and cannot have gaps. Thus, $M$ is even.  We
take each pulse to have a duration $1/2M$, without losing generality.

\begin{figure}
\includegraphics[width=8cm]{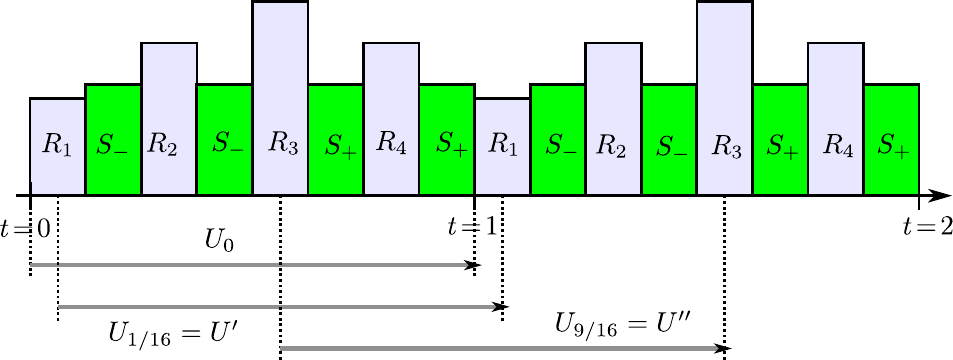}
\caption{(Color online) A DTQW is defined by a periodic sequence of
  pulses: site-dependent spin rotations $R_j$,
  Eq.~\eqref{eq:r_chi_theta_def}, and spin-dependent translations
  $S_+$ and $S_-$, Eq.~\eqref{eq:s_pm_def}.  The unitary timestep
  operator $U_0$ corresponds to a complete period, as in
  Eq.~\eqref{eq:U_def}. The same quantum walk can also be described in
  different ``timeframes'', i.e., by time-shifted timestep operators
  $U_T$ as in Eq.~\eqref{eq:UT_def}. Two examples are shown, with $U'$
  defined by $T=1/16$ and $U''$ by $T=9/16$.  }
\label{fig:pulses}
\end{figure}

We can also shift the starting time of the period by $T$, giving the
dynamics as $\ket{\Psi(t+1+T)} = U_T \ket{\Psi(t+T)}$, for any $t\in
\mathbb{Z}$. We refer to this shift, illustrated in Fig.~\ref{fig:pulses} as going into the ``timeframe''
$T$. The starting time $T$ has to be during a rotation, since
performing only part of a shift operation would leave the walker
between sites, and its description would necessitate an increased
Hilbert space.  This restricts $T$ to $T=(l-1)/M + y/(2M)$, for $1\le
l<M$, and $0\le y <1$. The Floquet operator in the timeframe $T$ reads
\begin{align}
U_T &= R_{M+l}^y S_{M+l-1} R_{M+l-1} \ldots S_{l+1} R_{l+1} S_l
R_l^{1-y},
\label{eq:UT_def}
\end{align}
where we define $R_j^{y} \equiv \sum_{x} \ket{x}\bra{x}\otimes
R(\chi_j(x),y\, \theta_j(x))$. Note that $U_T$ is a unitary transform
of $U_0$. 

A DTQW can be seen as a stroboscopic simulator of an effective
Hamiltonian $\HeffT$. The effective Hamiltonian is associated to the
Floquet operator by
\begin{align}
U_T &\equiv e^{-i H_{\mathrm{eff},T}}.
\label{eq:Heff_def}
\end{align}
The effective Hamiltonian is uniquely defined by this equation if we
restrict its eigenvalues, the quasienergies, to an ``energy Brillouin
zone'', $-\pi < \varepsilon \le \pi$.  This is completely analogous to
the restriction of the quasimomentum to the first Brillouin zone.

Previous work on CS in DTQWs has focused on a single timeframe,
whether $H_{\text{eff},0}$ has CS, and identifying the associated
topological invariant. Our crucial insight is that it is important to
widen the scope: \emph{a DTQW has CS, if there is a timeframe where
  its effective Hamiltonian has CS}: If a time $T$ and a unitary
operator $\Gamma$ acting on the coin space can be found, with
$\Gamma^2=1$, that $\Gamma U_{T} \Gamma = U_T^{-1}$. 


A sufficient condition for a DTQW to have CS represented by $\Gamma =
\sigma_x$, is that the sequence of operations defining the walk has an
``inversion point''.  By this, we mean that there is an $l$, with
which for every $j$:
\begin{align}  
R_{l-j} &= R_{l+j};\\
S_{l-j} = S_+ &\leftrightarrow S_{l+j+1} = S_-.
\label{eq:inversion_point}
\end{align}  
We choose $\Gamma=\sigma_x$ for two reasons. First, 
\begin{align}
\sigma_x S_- \sigma_x = S_+^{-1},
\label{eq:sigma_s}
\end{align}
whereby also $\sigma_x S_+ \sigma_x = S_-^{-1}$.
Second, since the local unitaries are rotations $R_j$ about axes
that have no $x$-component,
\begin{align} 
\sigma_x R(\chi,\theta) \sigma_x &= R(\chi,\theta)^{-1} = R(\chi,-\theta).
\label{eq:sigma_r}
\end{align} 
Consider the sequences of $M$ operations just after and just before
the middle of the ``inversion point'',
\begin{align}
F &= R_{l+M/2}^{1/2} S_{l-1+M/2} R_{l-1+M/2} \ldots 
R_{l+1}S_{l} R_l^{1/2};\\
G &= R_{l}^{1/2} S_{l-1} R_{l-1} \ldots  
R_{l+1-M/2}S_{l-M/2} R_{l-M/2}^{1/2}.
\end{align}
These give us two Floquet 
operators for the walk:
\begin{align} 
U' &= FG;& U'' &= GF,
\end{align} 
as shown in Fig.~\ref{fig:pulses}.  Using relations \eqref{eq:sigma_s}
and \eqref{eq:sigma_r}, we have that time reversal can be done
\emph{during} a period, $\Gamma F \Gamma G = 1$, whereby $G = \Gamma
F^{-1} \Gamma$.
From this it is straightforward to show that both $U'$ and $U''$ are
chiral symmetric. Thus, ``inversion symmetry'' of the DTQW sequence in
the sense of eq.~\eqref{eq:inversion_point} gives two inequivalent
``CS timeframes'': timeframes where the effective Hamiltonian of the
DTQW has CS.



CS allows a definition of sublattices, via the projection
operators $\Pi_A=(1+\Gamma)/2$, $\Pi_B=(1-\Gamma)/2$. 
Eigenstates of $\Heff'$ with quasienergy $\varepsilon\neq0,\pi$ can be
chosen to have equal support on both sublattices. Stationary states
with quasienergies $0$ or $\pi$, however, can be chosen to be on a
single sublattice in a timeframe with a CS Floquet operator, $U'$
(their wavefunctions in this timeframe are eigenstates of
$\Gamma$).

We now proceed to derive the bulk--boundary correspondence for DTQW
with CS.  We consider an inhomogeneous DTQW with CS, consisting of a
translationally invariant ``$L$'' bulk at $1 \ll x \ll d$ and an
``$R$'' bulk at $d \ll x \ll N$. There are (smooth or sharp)
boundaries between the two bulks around $x\approx d$ and $x\approx
1$. In the timeframe where the Floquet operator $U'$ has CS, the two
bulks have effective Hamiltonians $H_{\text{eff},L}'$ and
$H_{\text{eff},R}'$. We assume both bulk Hamiltonians have gaps around
$\varepsilon=0$ and $\varepsilon=\pi$.  Therefore, if stationary
states with quasienergies $\varepsilon=0$ or $\pi$ exist, they must
have wavefunctions confined to the edges, exponentially decaying
towards the bulks. The number of edge states at the edge around
$x\approx d$ on sublattice $A$
($B$) is $m_A'$ ($m_B'$). These can further be written as 
\begin{align}
m_A'&=m_{A,0}'+m_{A,\pi}';&
m_B'&=m_{B,0}'+m_{B,\pi}',
\label{eq:mprime_def}
\end{align}
where the second index stands for the energy.  We are looking for the
topological invariants of the bulk parts of the walk,
$\nu_{L,\varepsilon}$, and $\nu_{R,\varepsilon}$, where
$\varepsilon=0,\pi$, whose differences give us the number of
topologically protected edge states separately for each energy,
\begin{align}
\nu_{L,\varepsilon}-\nu_{R,\varepsilon} &=
m_{A,\varepsilon}'-m_{B,\varepsilon}'. 
\label{eq:nu_required}
\end{align}

The first step towards the topological invariants is the standard
winding number $\nu'$ \cite{schnyder_tenfold} associated to the bulk
effective Hamiltonian $H_{\text{eff}}'$, in the timeframe where the
Floquet operator is $U'$. This is obtained from the bulk,
translational invariant part of the Floquet operator, diagonal in
momentum space: $U' = \sum_k \ket{k}\bra{k}\otimes U'(k)$, and
$U'(k)=e^{-iH_{\text{eff}}(k)}$. Instead of the effective
Hamiltonian, it is convenient to calculate with $H'(k) = \sin [
  H_{\text{eff}}(k)]$. This has the same CS and the same winding
number as $H_{\text{eff}}(k)$ (and as its flattened version
$Q=\text{sgn}[H_{\text{eff}}(k)]$ \cite{schnyder_tenfold}), but can
be obtained much more efficiently via
$H'(k)=[U'(k)^\dagger-U'(k)]/(2i)$.  In a basis where
$\Gamma=\text{diag}(1,\ldots,1,-1,\ldots,-1)$ is a diagonal matrix
with an equal number of $+1$ and $-1$ elements, the matrix of
$H'(k)$ is block off diagonal because of CS. We name its upper right
block $h'(k)$.  The winding number $\nu'$ reads
\begin{align}
\nu' &= \frac{1}{2\pi i} \int_{-\pi}^{\pi}dk \frac{d}{dk} \log \det
h'(k).
\label{eq:nu_winding_def}
\end{align}

The winding number $\nu'$ is related to the difference of the bulk
polarizations on the two sublattices in bulk
\cite{future}. Therefore, it cannot differentiate between
$0$ and $\pi$ energy edge states, and can only be used to obtain the
sum of all topologically protected edge states around $x\approx d$:
\begin{align}
\nu_L' - \nu_R' &= m_{A,0}'+m_{A,\pi}' - m_{B,0}' - m_{B,\pi}'.
\label{eq:edge_difference_prime}
\end{align}
However, there is the other CS timeframe, $U''$, where we have 
\begin{align}
\nu_L'' - \nu_R'' &= m_{A,0}''+m_{A,\pi}'' - m_{B,0}'' - m_{B,\pi}''.
\label{eq:edge_difference_2prime}
\end{align}
We need to combine the information from the two CS timeframes to obtain the
topological invariants.  

We can obtain a simple connection between the two CS timeframes by
considering an edge state. In the timeframe of $U'=FG$, the edge state has
a wavefunction $\Psi$, entirely on sublattice $A$ (or $B$), i.e.,
$\Gamma\Psi = (-1)^g \Psi$, with $g=0$ (or $1$).  In other timeframes,
where $U_T$ has no CS, the energy of the edge state has to remain the
same, but its wavefunction can extend over both sublattices. In the
other CS timeframe $U''=GF$, however, its wavefunction, $\Phi = G \Psi$,
again has to to be entirely on a single sublattice. This can be $A$
(or $B$), whereby $\Gamma\Phi = (-1)^f \Phi$, with $f=0$ (or
$1$). Consider $G F \Phi = G \Gamma G^{-1} \Gamma \Phi = G \Gamma
G^{-1} (-1)^f \Phi = G \Gamma (-1)^f \Psi= (-1)^{g+f} G \Psi =
(-1)^{g+f} \Phi$. This shows that $0$ ($\pi$) energy edge states are
on the same (opposite) sublattice in the two CS timeframes.  This can be
summarized as
\begin{align}
m_{A}'' = m_{A,0}'+ m_{B,\pi}'; \quad m_{B}'' = m_{B,0}'+ m_{A,\pi}'.
\label{eq:m2prime}
\end{align}

To obtain the number of protected edge states at zero and $\pi$
energies separately, we substitute Eqs.~\eqref{eq:m2prime} into
Eqs.~\eqref{eq:edge_difference_prime}, \eqref{eq:mprime_def} and
\eqref{eq:edge_difference_2prime}, and rearrange to obtain
\begin{align}
m_{A,0}'-m_{B,0}' &= \frac{\nu_L'+\nu_L''}{2} - \frac{\nu_R'+ \nu_R''}{2};\\
m_{A,\pi}'-m_{B,\pi}' &= \frac{\nu_L'-\nu_L''}{2} - \frac{\nu_R'- \nu_R''}{2}.
\label{def:Q0}
\end{align}
We compare this with Eq.~\eqref{eq:nu_required}, and read off
the bulk topological invariants $(\nu_0,\nu_\pi)$ as
\begin{align}
(\nu_0, \nu_\pi) &= \left(\frac{\nu'+\nu''}{2}, \frac{\nu'-\nu''}{2} \right).
\label{def:topinv}
\end{align}
This, the bulk--edge correspondence for DTQW's with CS, is the main
result of the paper.


Having derived a general formula for the topological invariant of 1D
DTQWs with CS, we now discuss an example where the differences between
CS and PHS come into play. To arrive to the example, first consider
the ``split-step walk'' of Kitagawa et al.\cite{kitagawa_exploring},
given by $U_0 = S_+ R(0,\phi) S_- R(0,\theta_1)$. There, both PHS and
CS are present, and we find that the $\nu_\varepsilon$ are one-to-one
functions of the invariants $Q_\varepsilon$ induced by PHS
\cite{asboth_prb}: $\nu_\varepsilon = 1/2 -Q_\varepsilon$, for both
$\varepsilon=0,\pi$. (An interesting special case is the simple
quantum walk, obtained by setting $\phi=0$.) We can break PHS by using
nonzero angles $\chi$. To be able to break CS, we consider a
longer period of pulses, a ``4-step DTQW'', given by
\begin{align}
U_0 &= S_+ R_4 S_+ R_3 S_- R_2 S_- R_1.
\label{eq:Floquet4_def}
\end{align}
This walk has no CS if $R_2\neq R_4$, but has CS if $R_2=R_4$, with
$F=R_3^{1/2} S_- R_2 S_- R_1^{1/2}$ and $G=R_1^{1/2} S_+ R_4 S_+
R_3^{1/2}$. The 4-step walk also has the advantage that the effective
Hamiltonian will have longer range hoppings, and thus we can expect
higher values of the winding numbers. This is entirely analogous to
adding a 3rd nearest neighbor hopping term to the SSH model.


\begin{figure}
\includegraphics[width=8cm]{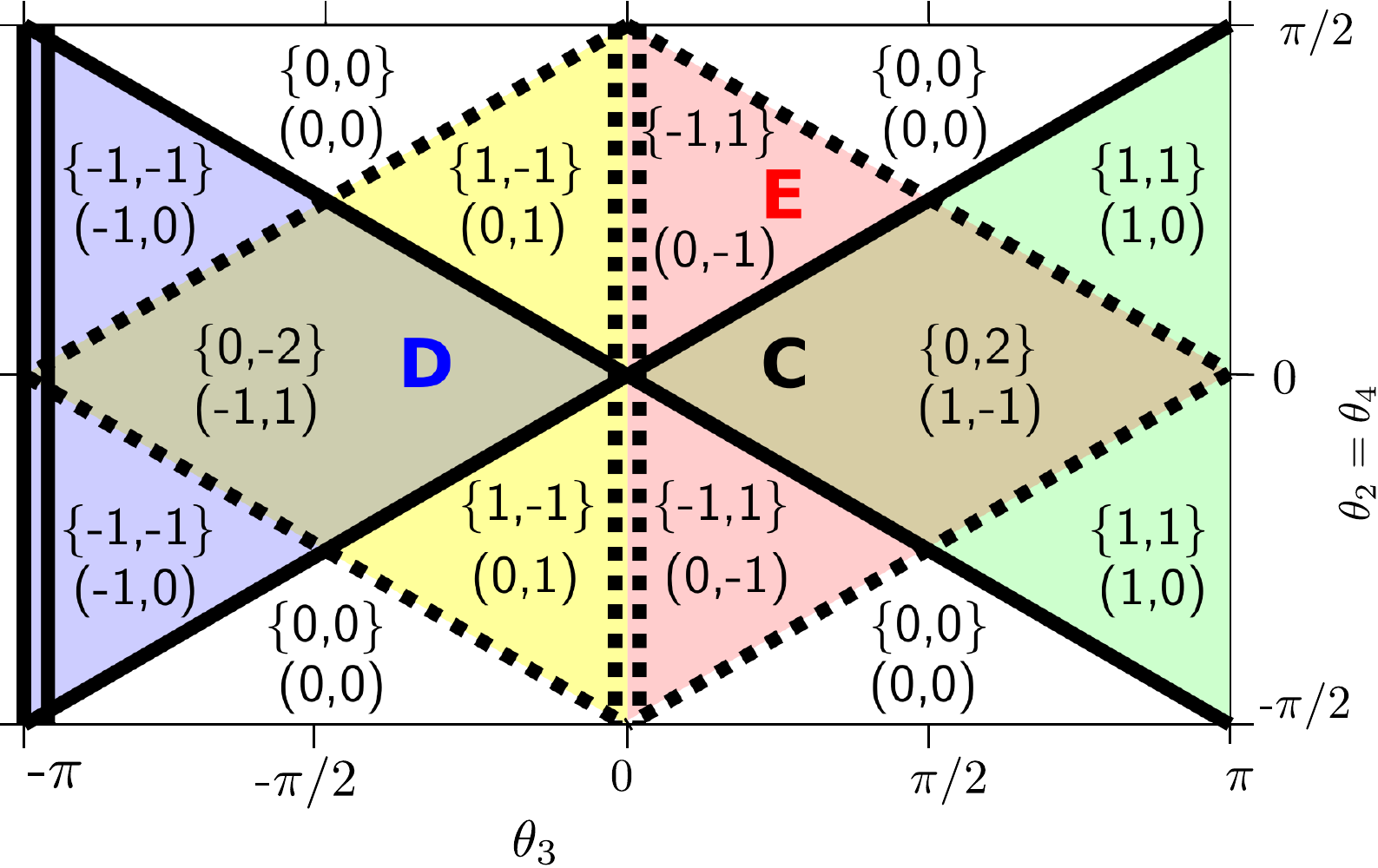}
\caption{(Color online) Parameter space of the 4-step DTQW with PHS ensured by
  $\chi_j=0$, CS ensured by $\theta_2=\theta_4$, and $\theta_1$ set to
  0. The DTQW has effective Hamiltonians with gaps around both
  $\varepsilon=0$ and $\varepsilon=\pi$, except at the gapless points
  where gaps close at $\varepsilon=0$ (solid lines) or
  $\varepsilon=\pi$ (dashed lines). Single lines indicate that the gap
  closes at a single $k$, at either $k=0$ or $k=\pi$. Double lines
  indicate double gap closings, at $k=\pm\pi/2$. For each gapped
  domain, the corresponding pair of winding numbers $\{\nu',\nu''\}$
  as well as the pair of topological invariants $(\nu_0,\nu_\pi)$,
  cf. Eq.~\eqref{def:topinv}, are shown. Letters ``C'', ``D'' and
  ``E'' indicate sets of parameters used for the inhomogeneous quantum
  walk, with rotation as in Eq.~\eqref{eq:R2_define}.}
\label{fig:map_invariants}
\end{figure}

The topological invariants in a section of the phase space of the
4-step DTQW with both CS and PHS (Cartan class BDI
\cite{schnyder_tenfold}) are shown in
Fig.~\ref{fig:map_invariants}. Here we set all $\chi_j=0$ to ensure
PHS, $\theta_2=\theta_4$ to ensure CS, and set $\theta_1=0$ for
simplicity.  We restrict $\theta_2$ to $-\pi/2<\theta_2<\pi/2$ since
adding $\pi$ to both $\theta_2$ and $\theta_4$ just brings two factors
of $-1$ that cancel out in both timeframes with CS. Generic values of
$\theta_2=\theta_4$ and $\theta_3$ give effective Hamiltonians with
gaps around both $\varepsilon=0$ and $\varepsilon=\pi$. Examples for
these are the points $C (\theta_2=\pi/20,\theta_3=\pi/4)$, $D$
($\theta_2=0, \theta_3 =-\pi/4$), and $E (\theta_2=\pi/4,
\theta_3=\pi/4)$.


To see the effects of breaking the symmetries on edge states, we
consider two inhomogeneous systems, consisting of two domains of 40
sites each, with sharp boundaries in between. The inhomogeneous
rotations read
\begin{align}  
R_{j} &= \sum_{x=1}^{40} \ket{x}\bra{x} \otimes R_{j,X}
+ \sum_{x=41}^{80} \ket{x}\bra{x} \otimes R_{j,C},
\label{eq:R2_define}
\end{align} 
where $X=D$ or $X=E$, and $C$ refer to the parameter sets of defined
in the previous paragraph and indicated in
Fig.~\ref{fig:map_invariants}.

\begin{figure}
\includegraphics[angle=270,width=\columnwidth]{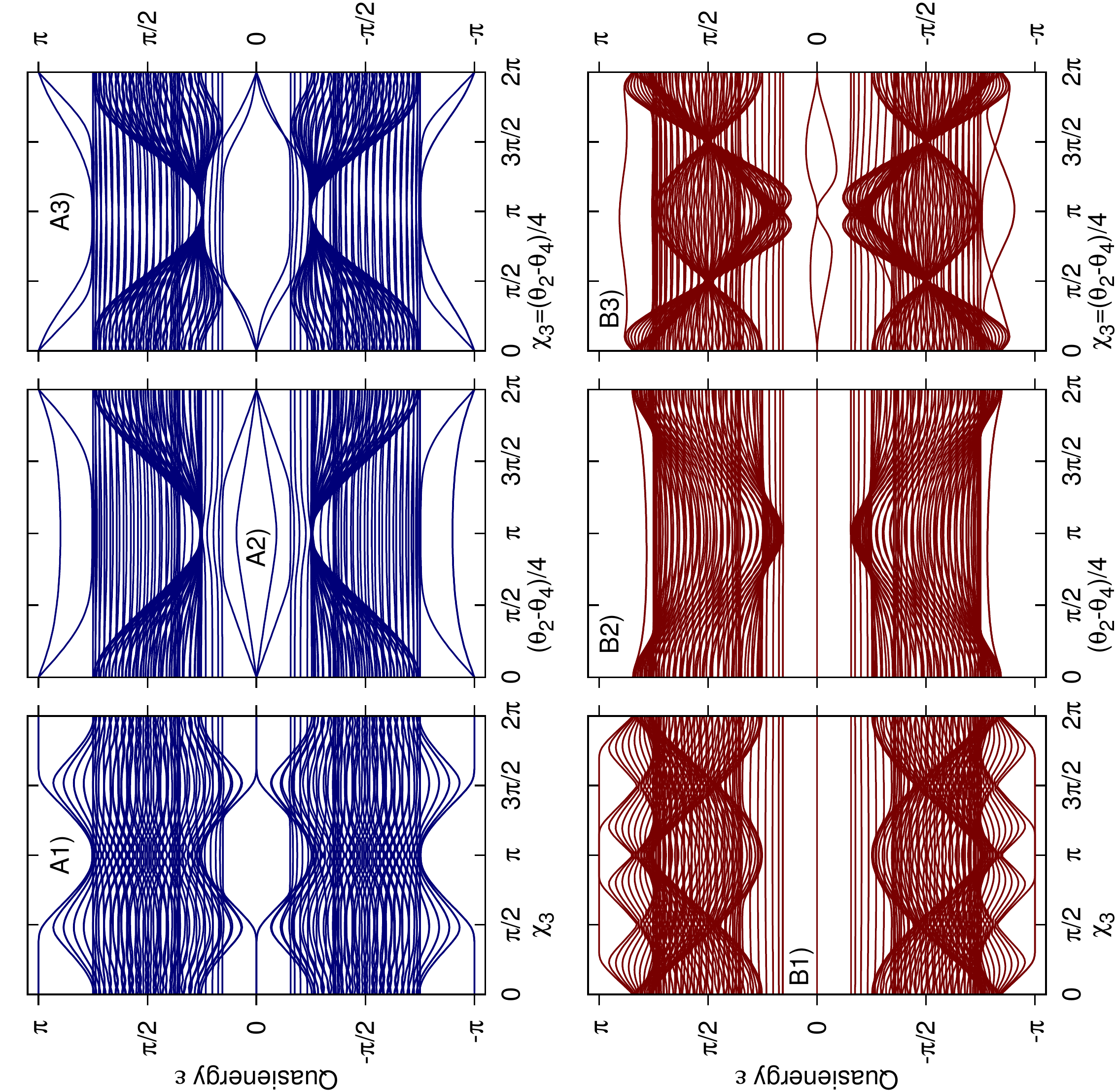}
\caption{(Color online) Spectra of an inhomogeneous ``4-step walk'' on $N=60$ sites
  as defined by Eqs.~\eqref{eq:Floquet4_def},\eqref{eq:R2_define}, with two
  domains: D and C (top row), and E and C (bottom row). Left panels:
  We break PHS via $\chi_3$ for $x<41$ on A1) and B1). As long as the
  bulk gaps are still open, the edge state energies do not change as
  they are still protected by CS. Middle panels, A2) and B2): we break
  CS by setting $\theta_2\to \theta_2+\Delta\theta$, $\theta_4\to
  \theta_4-\Delta\theta$ for $n<31$. This lifts the degeneracy of edge
  states on the same edge pairwise. At the interface between E and C,
  the unpaired edge state remains (B2). Right panels, A3) and B3): as
  both PHS and CS 
are
broken, no topologically protected edge states
  remain. In B3), the unpaired edge states at both edges are displaced
  in energy.  }
\label{fig:edges}
\end{figure}

We break PHS (realizing Cartan class AIII) in a controlled way by
introducing a nonzero $\chi_3$ in the bulk $0<x<41$. As long as the
bulk gaps are still open, breaking PHS does not change the edge state
energies, as shown in Fig.~\ref{fig:edges} A1), B1). The edge state
energies are still protected by CS, and can only move from their
original values if the bulk gap closes (at $\chi_3=\pi/2$ for the D-C
boundary).  We break CS (realizing Cartan class D), by changing
$\theta_2-\theta_4$ in the ``$L$'' bulk. A pair of edge states on the
same edge at the same energy can now break apart, becoming PHS
partners of each other. This can be seen in Fig.~\ref{fig:edges} A2)
at both $0$ and $\pi$ energy. However, a single edge state, as the one
between bulks $B$ and $C$, is still protected by PHS when CS is
broken, as seen in Fig.~\ref{fig:edges} B2).  Finally, to check that
no extra hidden symmetries remain, we break both CS and PHS (realizing
Cartan class A). In that case the edge state energies are not
protected anymore, cf Fig.~\ref{fig:edges} A3), B3). This shows that
our description of the relevant symmetries of the DTQW was indeed
exhaustive.

To summarize, we gave a definition of CS for DTQWs, and derived the
corresponding bulk topological invariants, using the fact that the
walk is defined by a sequence of operations, rather than just by its
unitary timestep operator.  The ``time-shifting'' approach presented
here based on finding the ``CS timeframes'' should generalize to
periodically driven quantum systems
\cite{floquet_topological,dora_review,weyl_metal_ol,kitagawa_periodic,4dquantum_1d}.
In such setups, PHS has been shown to lead to 2 types of ``Floquet
Majorana fermions'' \cite{akhmerov_majorana}, which should have clear
signals in transport \cite{transport_majorana_floquet} and can also be
useful for quantum information processing \cite{floquet_qubits}.
Theoretical proposals have already seen several such states at a
single edge if the driving also ensures CS \cite{many_majoranas}. The
bulk topological invariant controlling the number of these edge states
is as yet unknown, but it could be derived using the approach of this
paper.

Although 2-dimensional DTQWs have already been realized in experiments
\cite{schreiber_science}, their topological invariants are largely
unexplored. In 2 dimensions, edge states can exist in the absence of
symmetries; the related bulk--boundary correspondence for periodically
drive systems has only recently been found \cite{rudner_driven}. The
approach of identifying the ``symmetric timeframes'' could be a key idea
for the description of other symmetry classes for both periodically
driven quantum systems and DTQWs.

JKA thanks J.~Edge and A.~G\'abris, and HO thanks N.\ Kawakami,
Y.\ Nishimura, and T.\ Kitagawa for helpful discussions.  This
research was realized in the frames of TAMOP 4.2.4. A/1-11-1-2012-0001
''National Excellence Program -- Elaborating and operating an inland
student and researcher personal support system'', subsidized by the
European Union and co-financed by the European Social Fund.  This work
was also supported by the Hungarian National Office for Research and
Technology under the contract ERC\_HU\_09 OPTOMECH and the Hungarian
Academy of Sciences (Lend\"ulet Program, LP2011-016). 
H.\ O.\ 
was supported by Grant-in-Aid (No.\ 25800213 and No.\ 25390113) from the
Japan Society for Promotion of Science.

\bibliography{walkbib}{}

\end{document}